\documentstyle[aps]{revtex}

\begin{document}

\title{Theory of the Shubnikov-de Haas effect in quasi-two-dimensional metals}
\author{P.D. Grigoriev }
\address{
  Grenoble High Magnetic Field Laboratory, MPI-FKF and CNRS, \\
BP 166, F-38042 Grenoble Cedex 09, France. \\
  L.D.Landau Institute for Theoretical Physics, 142432
Chernogolovka, Russia.}
\date{\today }
\maketitle

\begin{abstract}
The Shubnikov - de Haas effect in quasi-two-dimensional normal
metals is studied. The interlayer conductivity is calculated
using the Kubo formula. The electron scattering on short-range is
considered in the self-consistent Born approximation. The result
obtained differs from that derived from the Boltzmann transport
equation. This difference is shown to be a general feature of
conductivity in magnetic field. A detailed description of the two
new qualitative effects -- the field-dependent phase shift of
beats and of the slow oscillations of conductivity is provided.
The results obtained are applicable to strongly anisotropic
organic metals and to other quasi-two-dimensional compounds.
\end{abstract}

\section{Introduction}

Magnetic quantum oscillations were discovered long ago and were
frequently used as a powerful tool of studying the geometry of
Fermi surfaces and other electronic properties of various metals.
In recent years, quasi-two-dimensional (quasi-2D) organic metals
\cite{OMRev} attract great interest because many new
unconventional effects are very pronounced in these compounds.
These effects are high-T$_{c}$ superconductivity, spin and charge
density waves, strong anisotropic magnetic quantum oscillations
etc. Much work was devoted to studying magnetic quantum
oscillations in these compounds (for a review see e.g. \cite
{MQORev}). The quantum oscillations of magnetization are a
thermodynamic effect that is completely determined by the
density-of-states distribution. Any exact calculation of the
electron density of states (DoS) is a very complicated problem but
a semi-phenomenological description of magnetization oscillations
in quasi-2D compounds was recently provided in a number of
theoretical papers \cite{dHvAth,Harrison,Pavel2}. The chemical
potential oscillations and the arbitrary electron reservoir due
to the open sheets of the Fermi surface create no principal
difficulties\cite {Pavel2}. Since the number of occupied LLs is
very large ($n_{F}>100$) in most quasi-2D organic metals, the
effect of the electron-electron interaction is reduced (as in the
Fermi liquid) and can be taken into account via the
renormalization of the electron effective parameters. Differences
in Landau level shape (which depends on a particular compound)
lead to only limited quantitative differences in magnetization
curves yielding no qualitatively new effects. On the qualitative
level, therefore, the de Haas-van Alphen (dHvA) effect in quasi-2D
normal metals is believed to be well understood.

Attempts of theoretical description of the quasi-2D Shubnikov-de
Haas (SdH) effect were not as successful although some work on
this subject appeared in recent years
\cite{Harrison,DS,Mc,Gv2,Av}. There are still many open
qualitative questions.

One of these open questions is the origin of the phase shift in
the beats of the resistivity oscillations with respect to those
in the magnetization. The beat behavior of the oscillations in
quasi-2D metals is known reliably to originate from a slight
warping of their Fermi surfaces in the direction normal to the 2D
plane. The superposition of the contributions from the maximum
and minimum cyclotron orbits leads to an amplitude modulation of
the $k$-th harmonic by the factor
$\cos (2\pi k\Delta F/2B-\pi /4)$, where $B$ is the magnetic field and $%
\Delta F=(c\hbar /2\pi e)(A_{max}-A_{min})$ is the difference
between the oscillation frequencies caused by the extreme orbits
with the $k$-space areas $A_{max}$ and $A_{min}$, respectively
\cite{Sh}. From the beat frequency one can readily evaluate the
warping of the Fermi surface and hence the interlayer transfer
integral $4t\approx \epsilon _{F}\Delta F/F$ (see e.g.
\cite{Wosnitza,SrRu}). The situation becomes less clear when the
warping is so weak that less than one half of the beat period can
be observed experimentally. In principle, an observation of one
single node would already be quite informative \cite{Weiss},
provided the phase offset (i.e. the phase of the beat at
$1/B\rightarrow 0$) is known. In the standard Lifshitz-Kosevich
(L-K) theory \cite{Abrik} this phase offset is strictly determined
by geometrical reasons and is equal to $-\pi /4$ for both the dHvA
and SdH effects \cite{Sh}.
However recent experiments on layered organic metals $\kappa $-(BEDT-TTF)$%
_{2}$ Cu[N(CN)$_{2}$]Br \cite{Weiss} and
(BEDT-TTF)$_{4}$[Ni(dto)$_{2}$] \cite{Balthes} revealed a
significant difference in the node positions of the beats of dHvA
and SdH signals. The phase shift in the latter compound was
estimated to be as big as $\pi /2$.

Another very interesting phenomenon (also not explained in the
framework of the standard theory) is slow oscillations of
magnetoresistance that were observed in a number of quasi-2D
organic metals \cite{MQORev,iBr2,tera,togo,broo}. The behavior of
slow oscillations resembles that of the SdH effect that lead to a
suggestion of additional, very small Fermi surface pockets in
these materials. However, band structure calculations (basically
giving a good description of the electron band structure and of
the Fermi surface topology in organic metals) show no evidence of
such small pockets in any of these compounds. Moreover, while
slow oscillations are often very pronounced in magnetoresistance,
sometimes even dominating the oscillation spectrum, thus far no
analogous observation in oscillating magnetization (dHvA effect)
was reported. Certainly, slow oscillations carry useful
information about the compounds but one needs some theoretical
explanation and, desirably, a quantitative description of the
phenomenon to extract this information. Note that in both cases
the oscillation spectrum was strongly dominated by the first
harmonic when no substantial deviations from the standard L-K
theory are expected.

 An explanation and the qualitative
description of phase shift of beats (which is based on the
Boltzmann transport equation) was proposed recently \cite{PhSh}
together with a comparative experimental study of this effect. An
idea that the slow oscillation may arise as an entanglement of
different rapidly oscillating contributing factors in
conductivity which have slowly oscillating amplitudes due to
beats was also recently suggested in \cite{SO} and proved by
presenting experimental results on temperature and angular
dependences of slow oscillations. In this paper we give a more
accurate theoretical description of these phenomena as well as a
detailed calculation of the interlayer conductivity in quasi-2D
metals in strong magnetic field.

In sec. II the general formula (\ref{Sigs}) is derived. The
calculation is performed starting from the Kubo formula. The
result of this calculation differs from that obtained using the
Boltzmann transport equation. The origin of this difference is
pointed out. The additional term is shown to be a general feature
of conductivity in magnetic field. However it becomes essential
only in the quasi-2D case. The simple explicit formula for
interlayer conductivity is obtained in the self-consistent Born
approximation in sec. III. A discussion of the reliability and
possible application of the results obtained is given in sec. IV.

\section{General formula for interlayer conductivity}

We consider a quasi-2D metal in magnetic field perpendicular to the
conducting layers: $\vec{B}\Vert \vec{z}$. The electron spectrum of quasi-2D
electron gas in magnetic field is then given by
\begin{equation}
\epsilon \left( n,k_{z}\right) =\hbar \omega _{c}\,(n+1/2)-2t\cos (k_{z}d)
\label{ES}
\end{equation}
where $t$ is the interlayer transfer integral, $k_{z}$ is the wavevector
perpendicular to the layers, $d$ is the interlayer distance, $\omega
_{c}=eB/m^{\ast }c$ is the cyclotron frequency. Both $\hbar \omega _{c}$ and
$t$ are assumed to be much smaller than the Fermi energy.

To calculate conductivity we use the Kubo formula \cite{Mah}. The procedure
is similar to that in three-dimensional metals without magnetic field (\cite
{Mah}, \S\ 7.1.2). In magnetic field only the new set of quantum numbers $%
m\equiv \{n,k_{z},k_{y}\}$ should be used instead of momentum $\vec{p}$ and
the alternative dispersion relation (\ref{ES}). The evaluation of the Kubo
formula without vertex corrections gives
\begin{equation}
\sigma _{zz}=\frac{e^{2}\hbar }{V}\sum_{m}v_{z}^{2}(m)\int \frac{d\epsilon }{2\pi }%
A^{2}(m,\epsilon )\left( -n_{F}^{\prime }(\epsilon )\right) \,  \label{Sigd1}
\end{equation}
where the volume $V$ normalizes the sum over quantum numbers $m$ , $e$ is
the electron charge, the limits of the integral over $\epsilon $ are $%
(-\infty ;\infty )$, $\ n_{F}^{\prime }(\epsilon )$ is the derivative of the
Fermi distribution function:
\begin{equation}
-n_{F}^{\prime }(\epsilon )=1/\{4T\cosh ^{2}\left[ (\epsilon -\mu )/2T\right]
\}  \label{nFd}
\end{equation}
and $A(m,\epsilon )$ is the spectral function that is related to the
electron Green's function $G^{R}(m,\epsilon )$ or to the retarded
self-energy part $\Sigma ^{R}(m,\epsilon ):$
\begin{equation}
A(m,\epsilon )\equiv -2\mbox{Im}G^{R}(m,\epsilon )=\frac{-2\mbox{Im}\Sigma
^{R}(m,\epsilon )}{\left[ \epsilon -\epsilon (m)-\mbox{Re}\Sigma
^{R}(m,\epsilon )\right] ^{2}+\left[ \mbox{Im}\Sigma ^{R}(m,\epsilon )\right]
^{2}}  \label{SFD}
\end{equation}
Formula (\ref{Sigd1}) is close to the corresponding formula
without magnetic field [\cite{Mah}, formula\ (7.1.10)] until the
self-energy part $\Sigma ^{R}(m,\epsilon )$ is specified. It
arises mainly from impurity scattering. The main contribution to
resistivity comes from the short-range impurity scattering. We
approximate short-range impurities by point-like ones. Then, if
one does not take into account the diagrams with intersections of
impurity lines in the self-energy (the contribution of such
diagrams at finite $k_z$ dispersion of electrons is usually
small) the electron self-energy depends only on electron energy
and not on electron quantum numbers. This fact greatly simplifies
the calculations because the sum over quantum numbers $m$ in
formula (\ref{Sigd1}) can be now computed analytically. The
constant part of the real part $\mbox{Re}\Sigma ^{R}(\epsilon )$
of electron self-energy produces only a constant shift of the
chemical potential. It does not influence the physical effects
and, hence, is omitted in the subsequent calculations. The small
oscillating part $\mbox{Re}{\tilde \Sigma ^{R}}(\epsilon )$ of
$\mbox{Re}\Sigma ^{R}(\epsilon )$ enters the final expression for
conductivity in the second order in small damping factors. Hence,
it can affect slow oscillations of conductivity and should be
kept in accurate quantitative analysis. It always come in the
combination $\epsilon ^* \equiv \epsilon -\mbox{Re}{\tilde \Sigma
^{R}}(\epsilon )$.

On the contrary, the imaginary part of self-energy $\mbox{Im}\Sigma
^{R}(\epsilon )$ is very important since it describes the momentum
relaxation of electrons.

Performing the summation over $k_{y}$ in (\ref{Sigd1}) and changing
integration over $k_{z}$\ by integration over energy $\epsilon (n,k_{z})$ we
get
\begin{eqnarray}
\sigma _{zz} &=&e^{2}\hbar N_{LL}\sum_{n}\int_{0}^{\pi
}\frac{d(k_{z}d)}{\pi } v_{z}^{2}(k_{z})\int \frac{d\epsilon
}{2\pi }A^{2}(\epsilon (k_{z},n),\epsilon )\,\left(
-n_{F}^{\prime }(\epsilon )\right) =  \nonumber
\\
&=&\,e^{2}N_{LL}d\int \frac{d\epsilon ^{\prime }}{\pi
}\sum_{n}\left| v_{z}(\epsilon ^{\prime },n)\right| \int
\frac{d\epsilon }{2\pi }  A^{2}(\epsilon ^{\prime },\epsilon
)\left( -n_{F}^{\prime }(\epsilon )\right) \label{Sigd2}
\end{eqnarray}
where $N_{LL} \equiv B/\Phi_0 d$ is the electron density on one
Landau level and the electron velocity $v_{z}(\epsilon ,n)$ is
given by (\ref{vz}). To go further we have to transform the sum
over LLs to a sum over harmonics. This can be done using the
Poisson summation formula (Appendix A). Substituting (\ref{Sv})
into (\ref{Sigd2}) we obtain:
\begin{eqnarray}
\sigma _{zz} &=&e^{2}N_{LL}\int \frac{d\epsilon ^{\prime }}{2\pi }
\sum_{k=-\infty }^{\infty }(-1)^{k}\frac{2td^{2}}{\hbar k}\exp
\left( \frac{ 2\pi ik\epsilon ^{\prime }}{\hbar \omega
_{c}}\right) J_{1}\left( \frac{4\pi kt}{\hbar \omega _{c}}\right)
\int \frac{d\epsilon }{2\pi }A^{2}(\epsilon ^{\prime },\epsilon
)\left( -n_{F}^{\prime }(\epsilon )\right) =  \nonumber
\\
&=&e^{2}N_{LL}\sum_{k=-\infty }^{\infty }(-1)^{k}\frac{2td^{2}}{\hbar k}%
J_{1}\left( \frac{4\pi kt}{\hbar \omega _{c}}\right) \int \frac{d\epsilon }{%
2\pi }\left( -n_{F}^{\prime }(\epsilon )\right) I_{z}(\epsilon ,k)
\label{Sigd2*}
\end{eqnarray}
where one should use the expansion $J_{1}(kx)/k=x/2$ for the
zeroth harmonic $k=0$, and the integral $I_{z}(\epsilon ,k)$ over $\epsilon
^{\prime }$ can be easily evaluated with the spectral function (\ref{SFD}):
\begin{eqnarray}
I_{z}(\epsilon ,k) &\equiv &\int \frac{d\epsilon ^{\prime }}{2\pi }%
A^{2}(\epsilon ^{\prime },\epsilon )\exp \left( \frac{2\pi ik\epsilon
^{\prime }}{\hbar \omega _{c}}\right) =  \label{Iz1} \\
&=&\int \frac{d\epsilon ^{\prime }}{2\pi }\left( \frac{-2\mbox{Im}\Sigma
^{R}(\epsilon )}{\left[ \epsilon ^* -\epsilon ^{\prime }\right] ^{2}+\left[ %
\mbox{Im}\Sigma ^{R}(\epsilon )\right] ^{2}}\right) ^{2}\exp \left( \frac{%
2\pi ik\epsilon ^{\prime }}{\hbar \omega _{c}}\right) =  \nonumber \\
&=&\exp \left( \frac{2\pi ik\,\epsilon ^*}{\hbar \omega _{c}}\right) \left(
\frac{1}{\left| \mbox{Im}\Sigma ^{R}(\epsilon )\right| }+\frac{2\pi k}{\hbar
\omega _{c}}\right) R_{D}(k,\epsilon )  \label{Iz}
\end{eqnarray}
where $\epsilon ^* \equiv \epsilon -\mbox{Re}{\tilde \Sigma
^{R}}(\epsilon )$ and
\begin{equation}
R_{D}(k,\epsilon )=\exp \left( -2\pi \left| k\right| \,\left| \mbox{Im}%
\Sigma ^{R}(\epsilon )\right| /\hbar \omega _{c}\right)  \label{RDSh}
\end{equation}
has the form similar to that of the usual Dingle factor $R_{D}(k)=\exp
\left( -2\pi ^{2}k\,k_{B}T_{D}/\hbar \omega _{c}\right) $\textbf{.}
Collecting formulas (\ref{Sigd2*}) and (\ref{Iz}) we get
\begin{eqnarray}
\sigma _{zz} &=&e^{2}N_{LL}\int \frac{d\epsilon }{2\pi }\left(
-n_{F}^{\prime }(\epsilon )\right) \sum_{k=-\infty }^{\infty }\frac{%
(-1)^{k}2td^{2}}{\hbar k}J_{1}\left( \frac{4\pi kt}{\hbar \omega _{c}}%
\right) \times  \nonumber \\
&&\times \exp \left( \frac{2\pi ik\,\epsilon ^* }{\hbar \omega _{c}}\right)
\left( \frac{1}{\left| \mbox{Im}\Sigma ^{R}(\epsilon )\right| }+\frac{2\pi k%
}{\hbar \omega _{c}}\right) R_{D}(k,\epsilon ).  \label{Sigs}
\end{eqnarray}

Note that this expression has additional term $2\pi k/\hbar
\omega _{c}$ near the standard $1/\left| \mbox{Im}\Sigma
^{R}(\epsilon )\right| $ term in round brackets in the second
line. This term can not be obtained from the Boltzmann transport
equation (compare, for example, with the results of \cite {PhSh}
and \cite{SO}) and it arises only due to quantization of electron
energy spectrum in magnetic field. This quantization results in
fast oscillations of the mean square electron velocity as
function of energy which give rise to the rapidly oscillating
factor $\exp \left( 2\pi ik\epsilon ^{\prime }/\hbar \omega
_{c}\right) $ in (\ref{Iz1}). This factor appears only in magnetic
field. The derivative of this factor with respect to $\epsilon
^{\prime }$ comes after the integration over $\epsilon ^{\prime }$
in (\ref{Iz1}) because the function $A^{2}(\epsilon ^{\prime
},\epsilon )=(G^{A} -G^{R})^2$ has one second-order pole in each
complex half-plane. These second-order poles arise from the
combinations $(G^{A})^2$ and $(G^{R})^2$, where $G^{A}$ and
$G^{R}$ are the advanced and retarded Green's functions
respectively. So, the additional term in (\ref{Sigs}) has quantum
origin; it appears due to fast oscillations of the mean squared
velocity at the Fermi level in magnetic field.

To go farther, we need an explicit form of the electron
self-energy which enters formula (\ref{Sigs}). It is calculated is
in the next section.

\section{Conductivity in self-consistent Born approximation}

\begin{figure}[htt]
\par
\begin{center}
\begin{picture}(120,55)
\put(0, 5){\Large{$\Sigma =$}}
\multiput(30,5)(0,14){3}{\line(0,1){10}}
\put(30, 46){\large{$\alpha $}}
\put(28, 41){x}
\put(30, 5){\circle*{3}}
\put(36, 5){\Large{+}}
\multiput(54,5)(7,14){3}{\line(1,2){5}}
\multiput(92,5)(-7,14){3}{\line(-1,2){5}}
\put(73, 46){\large{$\alpha $}}
\put(70, 41){\small{x}}
\put(54, 5){\circle*{3}}
\put(92, 5){\circle*{3}}
\put(54, 6){\line(1,0){38}}
\put(54, 4){\line(1,0){38}}
\end{picture}
\par
\end{center}
\caption{ The Dyson equation for the irreducible self-energy in
self-consistent Born approximation. The double solid line
symbolizes exact electron Green's function. } \label{SCBA}
\end{figure}
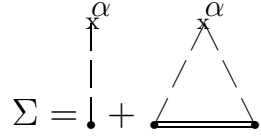

We consider electron scattering only by short-range impurities
because these impurities make the main contribution to the
relaxation of electron momentum. To calculate the electron
self-energy we use the self-consistent Born approximation. The
graphical representation of the Dyson equation
 for the irreducible self-energy part in
self-consistent Born approximation is shown in fig. \ref{SCBA}.
By such approximation we neglect multiple scattering on one
impurity (no more than two dash lines go to one impurity in fig.
\ref{SCBA}). The single dash line in fig. \ref{SCBA} corresponds
to the first-order term which leads only to a constant shift of
the chemical potential and, hence, can be omitted.

The corresponding analytical expression is
\begin{equation}
\Sigma ^{R}(m,\epsilon )=\left\langle \sum_{i}U^{2} \,
G(r_{i},r_{i},E)\right\rangle = C_i U^{2}\int d^3 r G(r,r,E)  \label{SER1}
\end{equation}
where $\sum_{i}$ is a sum over all impurities and the brackets
$\left\langle ..\right\rangle $ denote averaging over impurity
positions, $C_i$ is concentration of impurities which are assumed
to be uniformly distributed \cite{Comment1}. The electron Green's
function $G(r,r,E)$ in formula (\ref{SER1}) contains the
self-energy determined by the same formula (\ref{SER1}) (this is
why the approximation (\ref{SER1}) is called self-consistent Born
approximation). The Green's function is isotropic along the
conducting planes. Hence, one can write
\begin{equation}
G(r,r,E)= \left| \phi (z) \right| ^{2}G(E),  \label{G0d}
\end{equation}
where the electron wave function $\phi (z)$ along $z$-axis does not enter
the final result because it disappears after integration over $z$ in (\ref
{SER1}), and
\begin{equation}
G(E)=\frac{-N_{LL}}{\hbar \omega _{c}}\left\{ A(E)+i\pi \left[
1+2\sum_{k=1}^{\infty }(-1)^{k}J_{0}\left( \frac{4\pi kt}{\hbar \omega _{c}}%
\right) \exp \left( 2\pi ik\frac{E-\Sigma (E)}{\hbar \omega _{c}}\right)\,%
\right] \right\}.  \label{G0p}
\end{equation}
$A(E)$ is a slowly varying function of energy which can be taken at the
Fermi energy. The exact form of this function is not important for conductivity
in Born approximation.
Formula (\ref{G0p}) can be derived performing the
summation  over electron quantum numbers $m\equiv \{n,k_{z},k_{x}\}$ in the
definition of the Green's function:
\begin{equation}
G(r,r,E)=\sum_{n,k_{z},k_{x}}\frac{\Psi _{n,k_{z},k_{x}}^{\ast }(r)\Psi
_{n,k_{z},k_{x}}(r)}{E-\epsilon_{n,k_{z}}-\Sigma (E)}  \label{GSum}
\end{equation}
where $E_{n,k_{z},k_{x}}\ $is\ the electron energy in the state with quantum
numbers $m\equiv \{n,k_{z},k_{x}\}$ given by (\ref{ES}). The electron wave
function $\Psi_{n,k_{z},k_{x}}(r)$ in Landau gauge is approximately given by
\[
\Psi _{n,k_{z},k_{x}}(r)=\frac{e^{i(k_{x}x+k_{z}z)}}{\sqrt{L_{x}L_{z}}}\chi
_{n}(y-y_{0}) \phi (z)
\]
where $y_{0}=-c\hbar k_{x}/eB$ \ and the normalization condition
$\int_{-\infty }^{\infty }\left| \chi _{n}(y)\right| ^{2}dy=1$ is
used to perform integration over $k_{x}$. The further calculation
of the sum in (\ref{GSum}) is similar to that in (\ref{Sigd1}).

The Born approximation (formula (\ref{SER1})) takes into account
only the first term of expansion in the small parameter $\pi
UN_{LL}/\hbar \omega _{c}=\pi f/d$, where $N_{LL}/\hbar
\omega_{c}$ is equal to the electron density of states at the
Fermi level in unit volume, $f$ is the scattering amplitude
(which is constant at small wave vector $q\ll 1/r_{0}$, $r_{0}$
is the range of the impurity potential). For short-range
impurities the parameter $f/d$ is usually small.

From (\ref{SER1}) one can easily see that in Born approximation
the imaginary part of self-energy is proportional to the density
of states \cite{Comment3}:
\begin{equation}
-\mbox{Im}\Sigma ^{R}(\epsilon )=-C_{i}U^{2}\,\mbox{Im}G(\epsilon ) =\pi
\,C_{i}U^{2} \times \rho (\epsilon ).  \label{ImS}
\end{equation}
The unknown coefficient $C=\pi \,C_{i}U^{2}$ in (\ref{ImS}) is
simply related to the average Dingle temperature $T_{D}$:
$\left\langle \left| \mbox{Im}\Sigma ^{R}(m,\epsilon )\right|
\right\rangle =C\cdot \left\langle \rho (\epsilon )\right\rangle
=C\cdot \left( N_{LL}/\hbar \omega _{c}\right) (1+n_{R})=\pi
k_{B}T_{D} $, where the triangular brackets mean an average value
of a quantity inside, $k_{B}=1.38\cdot 10^{-16}erg/K$ is the
Boltzmann constant and $n_{R}$ is the density of reservoir states
that exist in many organic metals due to the open sheets of the
FS.

In the extreme 2D case $(\hbar \omega _{c}\gg t)$, substantial deviations
from formula (\ref{ImS}) are possible because the strong degeneracy of the
LLs makes the Born approximation not applicable. Since we consider the case $%
2t>\hbar \omega _{c}$ (when the beats of the oscillations exist)
and $f/d\ll 1$, we shall use (\ref{SER1}) for our subsequent
calculations that are now straightforward.

From the formulas (\ref{SER1}, \ref{G0d} and \ref{G0p}) we have
\begin{equation}
\left| \mbox{Im}\Sigma ^{R}(m,\epsilon )\right| =\pi k_{B}T_{D}\left(
1+2\sum_{k=1}^{\infty }(-1)^{k}J_{0}\left( \frac{4\pi kt}{\hbar \omega _{c}}%
\right) \cos \left( \frac{2\pi k\,\epsilon ^{\ast }}{\hbar \omega _{c}}%
\right) R_{D}(k,\epsilon )\right) .  \label{ImS1}
\end{equation}
Together with (\ref{RDSh}) this gives a nonlinear equation for $\mbox{Im}%
\Sigma ^{R}(m,\epsilon )$. We can solve it in the strong harmonic damping
limit using iteration procedure i.e. by making an expansion in the small
oscillating part which is an expansion in a parameter $\left( R_{D}\sqrt{%
\hbar \omega _{c}/2\pi ^{2}t}\right) .$ To treat the slow oscillation
accurately one has also to pick up all second-order slowly oscillating
terms. We get
\begin{equation}
\left| \mbox{Im}\Sigma ^{R}(\epsilon )\right| \approx \pi k_{B}T_{D}\Bigg\{%
1-2J_{0}\left( \frac{4\pi t}{\hbar \omega _{c}}\right) \cos \left( \frac{%
2\pi \,\epsilon }{\hbar \omega _{c}}\right) R_{0D} \Bigg\}.
\label{ImSf}
\end{equation}
where $R_{0D}=\exp \left( -2\pi ^{2}\,k_{B}T_{D}/\hbar \omega
_{c}\right) $. There is no slowly oscillating second-order term in
the self energy in Born approximation. At this point the real
part of the electron self energy was important because it
canceled the contribution from the entanglement with the
oscillations of the Dingle factor (\ref{RDSh}). In the second
order in damping factors any combination of the form
\[
\cos \left( \frac{2\pi \,(\epsilon -\mbox{Re} {\tilde \Sigma
^{R}} (\epsilon ))}{\hbar \omega _{c}}\right) \exp \left(
\frac{-2\pi \left| \mbox{Im}\Sigma ^{R}(\epsilon )\right| }{\hbar
\omega _{c}}\right) =\cos \left( \frac{2\pi \,\epsilon }{\hbar
\omega _{c}} \right) R_{0D}
\]
does not produce slowly oscillating term. This statement can be
easily checked by substituting (\ref{SER1}) with (\ref{G0p}) into
(\ref{ImS1}). If we neglected $\mbox{Re}\Sigma ^{R}(\epsilon )$
in (\ref{ImS1}) we would get an additional slowly oscillating term
\[
\Bigg\{-\frac{4\pi ^{2}\,k_{B}T_{D}}{\hbar \omega _{c}}J_{0}^{2}\left( \frac{%
4\pi t}{\hbar \omega _{c}}\right) R_{0D}^{2}\Bigg\}
\]
in curly brackets of (\ref{ImSf}) which arise from the mixing
with the oscillating Dingle factor and enters not only the
imaginary part of the self-energy but also the density of
electron states (see (\ref{ImSf})). The slow oscillations of
$\rho (\epsilon )$ would result in huge slow oscillations of
magnetization which are increased by an additional factor
$\epsilon _{F}/\hbar \omega _{c}$. Such huge slow oscillations of
magnetization have surely not been observed and the result
(\ref{ImSf}) is correct. Beyond the
Born approximation the relation (\ref{ImS}) is no more valid and $\mbox{Im}%
\Sigma ^{R}(\epsilon )$ acquires some slow oscillating term
although $\rho (\epsilon )$ does not. In the previous
(unpublished) work \cite{ShubE} the real part of electron
self-energy was disregarded and the entanglement with the
oscillations of Dingle temperature (producing additional large
contribution to the slow oscillations) was incorrectly taken into
account.

Substituting (\ref{ImSf}) into (\ref{Sigs}) we obtain the following
expression for the conductivity:
\[
\sigma _{zz}=e^{2}N_{LL}\int d\epsilon \left( -n_{F}^{\prime }(\epsilon
)\right) \frac{2t^{2}d^{2}}{\hbar ^{2}\omega _{c}\,\pi k_{B}T_{D}}\times
\]
\begin{eqnarray*}
\times  &&\Bigg\{\frac{\,1-\frac{\hbar \omega _{c}}{\pi t}J_{1}\left( \frac{%
4\pi t}{\hbar \omega _{c}}\right) \cos \left( \frac{2\pi \,\epsilon ^{\ast }%
}{\hbar \omega _{c}}\right) R_{D}(\epsilon )}{\left[ 1-2J_{0}\left( \frac{%
4\pi t}{\hbar \omega _{c}}\right) \cos \left( \frac{2\pi \,\epsilon ^{\ast }%
}{\hbar \omega _{c}}\right) R_{D}(\epsilon )\right] }-\\
&&-\frac{2\pi k_{B}T_{D}}{t}J_{1}\left( \frac{4\pi t}{\hbar \omega _{c}}%
\right) \cos \left( \frac{2\pi \,\epsilon ^{\ast }}{\hbar \omega _{c}}%
\right) R_{D}(\epsilon )\Bigg\}.
\end{eqnarray*}

If the transfer integral is large enough, $4\pi t>\hbar \omega _{c},$ one
can use the expansions of the Bessel function at large value of argument:
\begin{eqnarray}
J_{0}(x) &\approx &\sqrt{2/\pi x}\cos \left( x-\pi /4\right) \,,\,x\gg 1
\label{BFE} \\
J_{1}(x) &\approx &\sqrt{2/\pi x}\sin \left( x-\pi /4\right) \,,\,x\gg 1\,.
\nonumber
\end{eqnarray}
Then performing again expansion in small parameter $\left( R_{D}\sqrt{\hbar
\omega _{c}/2\pi ^{2}t}\right) $ and making use of the standard
trigonometric formulas we get
\begin{eqnarray}
&&\sigma _{zz}=\frac{e^{2}N_{LL}2t^{2}d^{2}}{\hbar ^{2}\omega _{c}\pi
k_{B}T_{D}}\Bigg\{\,1+2\sqrt{\frac{\hbar \omega _{c}\left( 1+a^{2}\right) }{%
2\pi ^{2}t}}\cos \left( \frac{2\pi \,\mu }{\hbar \omega _{c}}\right) \times
\nonumber \\
&&\times \cos \left( \frac{4\pi t}{\hbar \omega _{c}}-\frac{\pi }{4}+\phi
_{b}\right) R_{D}R_{T}\,+  \label{Sigf} \\
&&+\frac{\hbar \omega _{c}}{2\pi ^{2}t}R_{D^{\ast } }^{2}
\sqrt{1+a_{S}^{2}}\cos \left[ 2\left( \frac{4\pi t}{\hbar \omega
_{c}}-\frac{\pi }{4}+\phi _{S}\right) \right] \Bigg\} \nonumber
\end{eqnarray}
where the phase shift of beats is
\begin{equation}
\phi _{b}=\arctan \left( a\right) ; \ a=\frac{\hbar \omega _{c}}{%
2\pi t}\left( 1+\frac{2\pi ^{2}k_{B}T_{D}}{\hbar \omega
_{c}}\right) \label{aph2}
\end{equation}
and the phase of slow oscillations is
\begin{equation}
\phi _{S}=\arctan \left( a_{S}\right) /2 \mbox{ \ \ where \ \
}a_{S}=\hbar \omega _{c}/2\pi t . \label{phs}
\end{equation}
The star index in the Dingle factor $R_{D^{\ast } }$ of slow
oscillations is explained in the next section. The temperature
smearing factor is given by the usual L-K expression:
\[
R_{T}=\frac{2\pi ^{2}k_{B}T/\hbar \omega _{c}}{\sinh \left( 2\pi
^{2}k_{B}T/\hbar \omega _{c}\right) }.
\]
It appear in the fast Shubnikov oscillations after integration
over energy of a rapidly oscillating function of energy with the
Fermi distribution function. The slowly oscillating term depends
only on the transfer integral $t$ and is independent of energy.
Hence, it does not acquire any temperature smearing.

The phase shift (\ref{aph2}) obtained from the Kubo formula is
larger than that of [\cite{PhSh}, formula (9)] obtained using the
Boltzmann transport equation by a factor $\left( 1+2\pi
^{2}k_{B}T_{D}/\hbar \omega _{c}\right) .$ This difference comes
from the additional term $2\pi k/\hbar \omega _{c}$ near
$1/\left| \mbox{Im}\Sigma ^{R}(\epsilon )\right| $ in round
brackets in the second line of (\ref{Sigs}). As has been noted
after formula (\ref{Sigs}), this term has quantum origin.
However, the result (\ref{Sigf}) and (\ref{phs}) concerning the
slow oscillation does not differ from [\cite{SO}, formula (4)].
Thus the Boltzmann transport equation and the Kubo formula in
self-consistent Born approximation give the same amplitude and
phase of slow oscillations.

\section{Discussion of the results}

In this paper a detailed calculation of the interlayer
magnetotransport in quasi-2D normal metals is performed. The
specific features of quasi-two-dimensionality and strong magnetic
field result in pronounced qualitative effects such as the phase
shift of beats of conductivity oscillations and the slow
oscillations which cannot be described in the framework of the
standard three-dimensional theory usually applied to quasi-2D
compounds. The beats of magnetoresistance oscillations in layered
compounds are used for estimating the interlayer transfer
integral which strongly influences different electronic
properties of strongly anisotropic compounds. The field-dependent
phase shift of beats may lead to the errors in this estimate.
Hence, a detailed quantitative description of this phenomenon is
important.

The result of the calculation using the Kubo formula is different
from that obtained using Boltzmann transport equation. An
additional term in conductivity (see formula (\ref{Sigs}) and the
discussion after this formula) is general for conductivity in
magnetic field. It arises due to Landau quantization of electron
energy spectrum. However, this term is proportional to the
oscillations of the electron mean square velocity which are
smaller than the oscillations of the electron relaxation time (or
of $-\mbox{Im}\Sigma ^{R}(\epsilon )\sim \rho (\epsilon )$) by the
same factor $\hbar \omega _{c}/2\pi t $ as the phase shift of
beats is. Therefore, this additional term in 3D case is smaller
than the main oscillating term by the factor of $\hbar \omega
_{c}/\epsilon_F$.

The slow oscillations of magnetoresistance can give useful
information about compounds under study. A significant feature of
the slow oscillations is that their Dingle factor $R_{D\ast }$ is
different from the factor $R_{D}$ of the Shubnikov oscillations.
The usual Dingle factor includes all temperature-independent
mechanisms of smearing of fast quantum oscillations. These are not
only microscopical scattering events of electrons but also
macroscopic spatial inhomogeneities of the sample. These
inhomogeneities lead to macroscopic spatial variations of the
electron energy $\epsilon ^* $ in formula (\ref{Sigs}) which is
equivalent to a local shift of the chemical potential. The total
signal is an average over the entire sample and such macroscopic
inhomogeneities lead to the damping of magnetic quantum
oscillations similar to that caused by temperature. Since the
slow oscillations do not depend on $\mu$, they are not affected
by this type of smearing and the corresponding Dingle temperature
$T_D^{\ast }$ of slow oscillations is determined by only
short-range scatterers. One can therefore estimate relative
contributions from macroscopic inhomogeneities and from local
defects to the scattering rate by comparing $T_D$ and $T_D^{\ast
}$. This role could be quite essential in organic metals. For
example, such a comparison for a sample of $\beta
$-(BEDT-TTF)$_2$IBr$_2$ gives \cite{SO} $T_D=(0.8\pm0.02)$ K
while $T_D^{\ast}=(0.15\pm0.02)$ K, that means that the long-range
crystal imperfections are important for damping of fast quantum
oscillations. The relaxation of electron momentum affecting
transport quantities comes, however, mainly from short-range
impurities.

The slow oscillations in formula (\ref{Sigf}) do not have
temperature damping factor. Hence, although the amplitude of the
slow oscillations contains the square of the Dingle factor (they
are a second-order effect), it can be larger than the amplitude
of the fast SdH oscillations at $T\gtrsim T_{D}$. However, our
experience disagrees with the statement that the slow
oscillations do not manifest any temperature dependence (and,
hence, could be seen at room temperature). Actually they do have
some temperature damping because the oscillating DoS itself has
some temperature dependence. The temperature damping of the DoS
oscillations comes from the electron-phonon and electron-electron
interactions. In normal 3D metals \cite {Abrik} the
electron-electron (e-e) scattering rate $1/\tau _{ee}\sim \left(
k_{B}T\right) ^{2}/\hbar \mu $ while the electron-phonon scattering rate $%
1/\tau _{ph}\sim \left( k_{B}T/\hbar \right) (k_{B}T/\hbar \omega
_{D})^{2}.$ One can estimate the effect of these scattering
processes on the DoS oscillations by introducing the additional
damping factor
\begin{equation}
R_{TD}\approx \exp \left[ -\pi (1/\omega _{c}\tau _{ee}+1/\omega
_{c}\tau _{ph})\right]  \label{RTD}
\end{equation}
analogous to the usual Dingle factor. This factor enters squared
in the amplitude of slow oscillations. The temperature $T_{SO}$ at
which the slow oscillations become damped by this factor is much
higher than the characteristic temperature of the damping of fast
quantum oscillations. It is approximately given by $\pi (1/\omega
_{c}\tau _{ee}(T_{SO})+1/\omega _{c}\tau _{ph}(T_{SO}))\approx 1$.
The above analysis of the temperature dependence of slow
oscillations is very approximate. A rigorous calculation must be
based on the exact calculation of the electron self-energy due to
these two types of interactions. Nevertheless, the above
arguments can give qualitative estimates. A more accurate
calculation would be useful since the temperature dependence of
slow oscillations at high enough temperature may give additional
information about the electron-phonon and electron-electron
interactions in various compounds where slow oscillations exist.
This is important since these interactions in layered organic
metals determine the superconducting and the density-wave
transitions.

The entanglement with the oscillations of chemical potential
contributes an additional temperature-dependent term to the slow
oscillations of conductivity. This term can be easily obtained by
substituting (\ref{muh}) into (\ref{Sigf}). However, this term
has additional damping factors $R_T ^2$ and $(R_{D}/R_{D^{\ast }
})^2$compared to the main slowly oscillating term. Therefore,
this correction is as small as the second harmonic of Shubnikov
oscillations is, and we can neglect it.

The slow oscillations does not appear in magnetization because
there is no suitable entanglement of different oscillating
quantities in magnetization. The magnetization being a
thermodynamic quantity is completely determined by the electron
density of states. However, the density of states does not have
slowly oscillating terms. The mixing with the oscillations of the
chemical potential, or with those of the Dingle factor and of
$\mbox{Re}\Sigma ^{R}(\epsilon )$ does not also lead to slow
oscillations of magnetization (see Appendix B).

Now we shall discuss the approximations made during the derivation
of formula (\ref{Sigf}).

The first limitation of the proposed analysis is that the
magnetic field is taken to be perpendicular to the conducting
layers. A finite tilt angle $ \theta $ of the magnetic field with
respect to the normal to the conducting planes may be
approximately taken into account by rescaling the Landau level
separation, $\omega _{c}\rightarrow \omega _{c}\cos \theta ,$ and
of the warping of the Fermi surface\cite{Yam}, $t(\theta
)=t(0)\,J_{0}(k_{F}d\tan \theta )$, where $k_{F}$ is the in-plane
Fermi momentum. But this is only a semiclassical approximation
based on the assumption that the FS remains the same. Actually,
the tilting of the magnetic field changes the dispersion relation
and a more profound study of the effect of tilting of magnetic
field on transport properties is required. The quantum mechanical
calculation of the dispersion relation in tilted magnetic field
in the first order of the transfer integral gives \cite{Kur}
$t(\theta )/t(0)=\exp \left( -g^{2}/4\right) L_{n}^{0}\left(
g^{2}/2 \right) $, where $g\equiv d\tan \theta /a_{H}$,
$a_{H}=\sqrt{\hbar c/eB_{z}}$ is the magnetic length and
$L_{n}^{0}\left( x\right) $ is Laguerre polynomial. This result
is also approximate, but it should work satisfactory at not too
great tilt angles. In the limit $n\rightarrow \infty $ the two
above results coincide.

More essential errors may come from the so-called incoherent or
weakly incoherent electron interlayer transport. This means, that
when the interlayer transfer integral becomes comparable to the
Dingle temperature, the electron interlayer jumps an electron
scattering on impurities should not be considered separately.
Their entanglement may result even in qualitatively new phenomena
in the quasi-2D magnetotransport \cite{Mc}. In the present work we
consider only the case $t > T_D$.

Other errors may come from the approximate expression for
self-energy (\ref{ImS}). The self-consistent Born approximation
at finite $k_z$ dispersion works quite well, but other scattering
mechanisms (especially for the calculation of the DoS) should be
taken into account. An accurate study of this problem may depend
on a particular type of the compound in hand. It may lead to some
quantitative modifications of formula (\ref{Sigf}).

The above analysis does not take into account the vertex
corrections. In our case (of point-like impurity scattering) this
is right because, according to the Ward identity, the vertex
$\vec{ \Gamma}(m,E)=\vec{p}+m\,\vec{\nabla}_{p}\Sigma ^{R}(m,E)$.
Hence, if the retarded self-energy depends only on energy, the
vertex corrections are zero. The fact that $\Sigma
^{R}(m,\epsilon )$ is approximately a function of energy
$\epsilon $ only is a consequence of the short-range (or
point-like) impurity potential. More precisely, if one takes a
point-like impurity potential and neglects all diagrams with the
intersections of the impurity lines in the self-energy, then
after averaging over randomly and uniformly distributed impurity
positions one obtains $\Sigma ^{R}(m,\epsilon )=\Sigma
^{R}(\epsilon )$. The neglected graphs with the intersections of
the impurity lines describe the coherent scattering on two
impurities simultaneously. The contribution of such scattering is
small at large enough interlayer transfer integral. In the
three-dimensional case without magnetic field the vertex
corrections produce an additional factor $(1-\cos \alpha )$ in
the integrand for the transport scattering relaxation time where
$\alpha $ is the scattering angle. But the scattering probability
is independent of the scattering angle in the case of point-like
impurities and the additive $\cos \alpha $ vanishes after the
integration over angles. Hence, the vertex corrections vanish.

In derivation of formula (\ref{Sigf}) only first- and
second-order terms in the small damping factors $R_T$ and $R_D$
were taken into account, assuming the harmonic damping to be
strong. This is valid in the most experiments on quasi-2D organic
metals where the amplitude of the second harmonic does not usually
exceed $5\% $ of the first harmonic.

So, in spite of the approximations made in the above analysis,
the proposed theoretical description is valid in a large domain
of parameters which one has in real experiments on quasi-2D
organic metals. For example, above result concerning the slow
oscillations and the phase of beats of the SdH oscillations is
important for an experimental study of these effects in $\beta
$-(BEDT-TTF)$_{2}$IBr$ _{2}$ (see, e.g. \cite{PhSh},\cite{SO}).
The proposed results may be also used for heterostructures with
large enough interlayer jumping.

The author thanks M.V. Kartsovnik, W. Biberacher, A.M. Dyugaev and I. Vagner
for encouragement and stimulating discussions. The work was supported by the
EU ICN contract HPRI-CT-1999-40013 and RFBR No. 00-02-17729a.

\appendix

\section{Transformation of the sums over LLs to the sums over harmonics}

To transform the sums over LL number into the harmonic sums we shall apply
the Poisson summation formula \cite{ZW}
\begin{equation}
\sum_{n=n_{0}}^{\infty }f(n)=\sum_{k=-\infty }^{\infty
}\int_{a}^{\infty }e^{2\pi ikn}f(n)\,dn
\end{equation}
 where $a\in (n_{0}-1;n_{0})$. This formula is valid for arbitrary function
$f(n).$ The electron velocity is determined from the dispersion
relation (\ref{ES}) as
\begin{eqnarray}
v_{z}(\epsilon ,n) &\equiv &\frac{\partial \epsilon (n,k_{z})}{\hbar
\partial k_{z}}=-\frac{2td}{\hbar }\sin (k_{z}d)=  \nonumber \\
&=&\frac{d}{\hbar }\sqrt{4t^{2}-\left( \epsilon -\hbar \omega
_{c}\,(n+1/2)\right) ^{2}}.  \label{vz}
\end{eqnarray}
The sum in (\ref{Sigd2}) now becomes
\begin{eqnarray}
\sum_{n}\,\left| v_{z}(\epsilon ,n)\right| &=&\sum_{n=0}^{\infty }\frac{d} {%
\hbar }\sqrt{4t^{2}-\left( \epsilon -\hbar \omega _{c}\left( \,n+\frac{1}{2}
\right) \right) ^{2}}=  \nonumber \\
&=&\frac{d}{\hbar }\hbar \omega _{c}\sum_{k=-\infty }^{\infty
}\int_{0}^{\infty }dn\, e^{ 2\pi ik\left( n-\frac{1}{2}\right) }
\,\sqrt{\left( \frac{2t}{\hbar \omega _{c}}\right) ^{2}-\left(
\frac{
\epsilon }{\hbar \omega _{c}}-\,n\right) ^{2}}  \nonumber \\
&=&\frac{d}{\hbar }\hbar \omega _{c}\sum_{k=-\infty }^{\infty
}\left( -1\right) ^{k}\exp \left( \frac{2\pi ik\epsilon }{\hbar
\omega _{c}}\right) \int_{-\infty }^{\infty }dx\, e^{ 2\pi ikx }
\,\sqrt{\left(
\frac{2t}{\hbar \omega _{c}}\right) ^{2}-x^{2}}  \nonumber \\
&=&\sum_{k=-\infty }^{\infty }\frac{dt}{\hbar } \frac{ \left(
-1\right) ^{k}}{k}\exp \left( \frac{2\pi ik\epsilon }{\hbar \omega
_{c}}\right) J_{1}\left( \frac{ 4\pi kt}{\hbar \omega
_{c}}\right)  \label{Sv}
\end{eqnarray}
In this formula for the zeroth harmonic $k=0$ one should use the
expansion $J_{1}(kx)/k=x/2$.

\section{Magnetization}

The first harmonic of the oscillating part of magnetization is
given by (see \cite{Pavel2}, formula 6)
\begin{equation}
\tilde{M}(B)=\frac{2N_{LL}\varepsilon _{F}}{\pi B}\sin \left(
\frac{2\pi \,(\varepsilon _{F}+\tilde{\mu}(B))}{\hbar \omega
_{c}}\right) J_{0}\left( \frac{4\pi t}{\hbar \omega _{c}}\right)
\,R_{T}R_{S}R_{D}(\varepsilon _{F}).
 \label{Ma}
\end{equation}
where the oscillating part of the chemical potential is
(\cite{Pavel2}, formula 5)
\begin{equation}
\tilde{\mu}(B)=\frac{\hbar \omega _{c}}{\pi (1+n_{R}(\varepsilon
_{F}))}\times \sin \left( \frac{2\pi \,(\varepsilon
_{F}+\tilde{\mu}(B))}{\hbar \omega _{c}}\right) J_{0}\left(
\frac{4\pi t }{\hbar \omega _{c}}\right) \,R_{T}R_{S}R_{D}.
 \label{muh}
\end{equation}
The entanglement of magnetization oscillations with the
oscillations of the Dingle factor (\ref{RDSh}) produces an
additional term
$$
\propto \sin \left( \frac{2\pi \varepsilon _{F}}{\hbar \omega
_{c}}\right) J_{0}\left( \frac{4\pi t }{\hbar \omega _{c}}\right)
\times \cos \left( \frac{2\pi \varepsilon _{F}}{\hbar \omega
_{c}}\right) J_{0}\left( \frac{4\pi t }{\hbar \omega _{c}}\right)
$$
$$
=\frac{1}{2}\sin \left( \frac{4\pi \varepsilon _{F}}{\hbar \omega
_{c}}\right) J_{0}^2 \left( \frac{4\pi t }{\hbar \omega
_{c}}\right)
$$ which give rise to the second harmonic but makes zero
contribution to the slow oscillations of magnetization.

The entanglement with the oscillations of the chemical potential
(\ref{muh}) produces the term
$$
\propto \sin \left( \frac{2\pi \,(\varepsilon
_{F}+\tilde{\mu}(B))}{\hbar \omega _{c}}\right) - \sin \left(
\frac{2\pi \,\varepsilon _{F}}{\hbar \omega _{c}}\right)$$
$$ = \sin \left( \frac{2\pi
\varepsilon _{F}}{\hbar \omega _{c}}\right) \left[ \cos \left(
\frac{2\pi \tilde{\mu}(B)}{\hbar \omega _{c}}\right) -1 \right] +
\cos \left( \frac{2\pi \varepsilon _{F}}{\hbar \omega
_{c}}\right) \sin \left( \frac{2\pi \varepsilon _{F}}{\hbar
\omega _{c}}\right) J_{0} \left( \frac{4\pi t }{\hbar \omega
_{c}}\right) \frac{2R_T R_D}{1+n_R}
$$
which also contribute only to the second harmonics (or higher
harmonics) but not to slow oscillations.


\begin{thebibliography}{99}
\bibitem{Comment1} In real layered compounds the impurity distribution
may have periodic modulation in the direction perpendicular to the layers
and with period equal to the interlayer distance. However, in Born approximation
this leads only to a constant
factor in (\ref{SER1}) which renormalizes the Dingle temperature.
\bibitem{Comment3} This fact has evident physical meaning. The matrix element of
scattering by a point-like impurity $U$ is the same for all
transitions not changing electron energy. Hence, the scattering
rate $1/\tau (\epsilon )=-2\mbox{Im}\Sigma ^{R}/\hbar $ is
proportional to the number of states to which an electron may
scatter, that is, to the density of states at a given energy.
\bibitem{Comment2} This fact
can be understood in a different way: slow oscillations originate
from local electron motion and are not sensitive to global
variations of the position of the chemical potential with respect
to the bottom of the conducting band.
\end{thebibliography}
\end{document}